\documentclass[preprint,5p]{elsarticle}
\usepackage{graphicx,amssymb,amsmath,amstext,hyperref,graphics,float,eurosym,geometry,lineno,changebar}
\nochangebars
\biboptions{sort&compress}
\hypersetup{
    colorlinks=true,        
    linkcolor=black,         
    citecolor=black,        
    filecolor=black,      
    urlcolor=black           
}

\newcommand{\fig}[1]{Fig.~\ref{#1}}

\begin{document}
\title{Modular transmission line probes for microfluidic nuclear magnetic resonance spectroscopy and imaging}
\author[1]{Manvendra Sharma}
\author[1]{Marcel Utz}
\address[1]{School of Chemistry, University of Southampton, Southampton SO17 1BJ, United Kingdom}
\begin{abstract} Microfluidic NMR spectroscopy can probe chemical and
bio-chemical processes non-invasively in a tightly controlled environment. We
present a dual-channel modular probe assembly for  high efficiency microfluidic
NMR spectroscopy and imaging. It is compatible with a wide range of microfluidic
devices,  without constraining the fluidic design. It collects NMR signals from
a designated sample volume on the device with high sensitivity and resolution.
Modular design allows adapting the detector geometry to different experimental
conditions with minimal cost, by using the same probe base. The complete probe
can be built from easily available parts. The probe body mainly consists of
prefabricated aluminium profiles, while the probe circuit and detector are made
from printed circuit boards. We demonstrate a double resonance HX probe with a
limit of detection of 1.4 nmol s$^{-1/2}$ for protons at 600~MHz, resolution of
3.35 Hz, and excellent B$_{1}$ homogeneity. We have successfully acquired
\textsuperscript{1}H-\textsuperscript{13}C and
\textsuperscript{1}H-\textsuperscript{15}N heteronuclear correlation spectra
(HSQC), including a \textsuperscript{1}H-\textsuperscript{15}N HSQC spectrum of
1 mM \textsuperscript{15}N labeled ubiquitin in 2.5 $\mu$l of sample volume.
\end{abstract} \maketitle
\section{Introduction}
\label{sec:intro}
NMR (Nuclear Magnetic Resonance) is a well established technique used in
chemical analysis~\cite{NMR-chemical-1995, hills1994magnetic,
rabenstein1991quantitative}, molecular structure
determination~\cite{wuthrich1990protein},
metabolomics~\cite{nmr-metabolomics-future-2017,nmr-metabolomics-2016}, reaction
monitoring~\cite{maiwald2004quantitative} etc. Microfluidic lab-on-a-chip (LOC)
devices are increasingly finding applications in chemistry and the life
sciences~\cite{whitesides2006origins,mark2010microfluidic}. They provide a
convenient way to integrate complex chemical and biochemical processes. In
particular, microfluidic technology allows detailed control over growth
conditions in the culture of cells, cell aggregates, and small
organisms~\cite{cellonchip-2006,cellonchip-review,tissueonchip-2008}.
Lab-on-a-chip devices can also enable high experimental throughput by
parallelisation. In spite of its analytical power, NMR has so far only rarely
been used in microfluidic systems. In general, information is extracted from
microfluidic devices only at the end point of the experiment. Typically, this is
done using fluorescent probes, which bind to one or a few analytes of interest.
This approach can reach very high sensitivity, with limits of detection down to
femtomolar concentrations. However, it is usually destructive, since the sample
is affected by the addition of the marker, and can only be done once, at the end
of the experiment. Following kinetic processes therefore requires using multiple
samples. By contrast, NMR spectroscopy is non-invasive, generic (virtually all
organic molecules give an NMR signal), and highly specific (no two metabolites
share the same NMR spectrum), and can therefore provide a useful complementary
readout capability for microfluidic devices. For example, the metabolic activity
of cell cultures can be followed non-invasively by NMR throughout the course of
the experiment~\cite{cellnmr-2015}, while effects on cell differentiation can
still be analysed by a (destructive) transcriptomic analysis at the end.
Unfortunately, the sensitivity of NMR is much lower than that of optical
techniques. The small volumes involved in microfluidic systems (typically in the
range of 1~$\mu$l) compound this problem. As is well known, the mass limit of
detection of inductive NMR detectors improves as they are scaled
down~\cite{Olson1995}. On this basis, a number of different micro-NMR detector
designs have been proposed and characterised in the
literature~\cite{utz2012review,micronmr2014review}. In most cases, these are
based on micro-solenoid~\cite{SUBRAMANIAN1998,Pines2007},
planar~\cite{Maguire2007,dieter2008,EHRMANN200}, or stripline
detectors~\cite{stripline_jan}. Significant research has gone into developing
micro-NMR detectors for flow probes\cbstart~\cite{bas-shim,Montinaro-2018,chen2017high,oosthoek2017continuous,bart2009microfluidic,bart2009optimization,massin2003planar}\cbend, where the sample container takes the form
of a fixed capillary into which the sample must be injected. Only a few designs
that allow insertion of a removable microfluidic device into the detector have \cbstart
been described~\cite{Spengler-2014,Spengler-2016,gream_2016,swyer2019digital}.\cbend

\begin{figure}
\centering
\includegraphics[width=.8\linewidth,keepaspectratio=true]{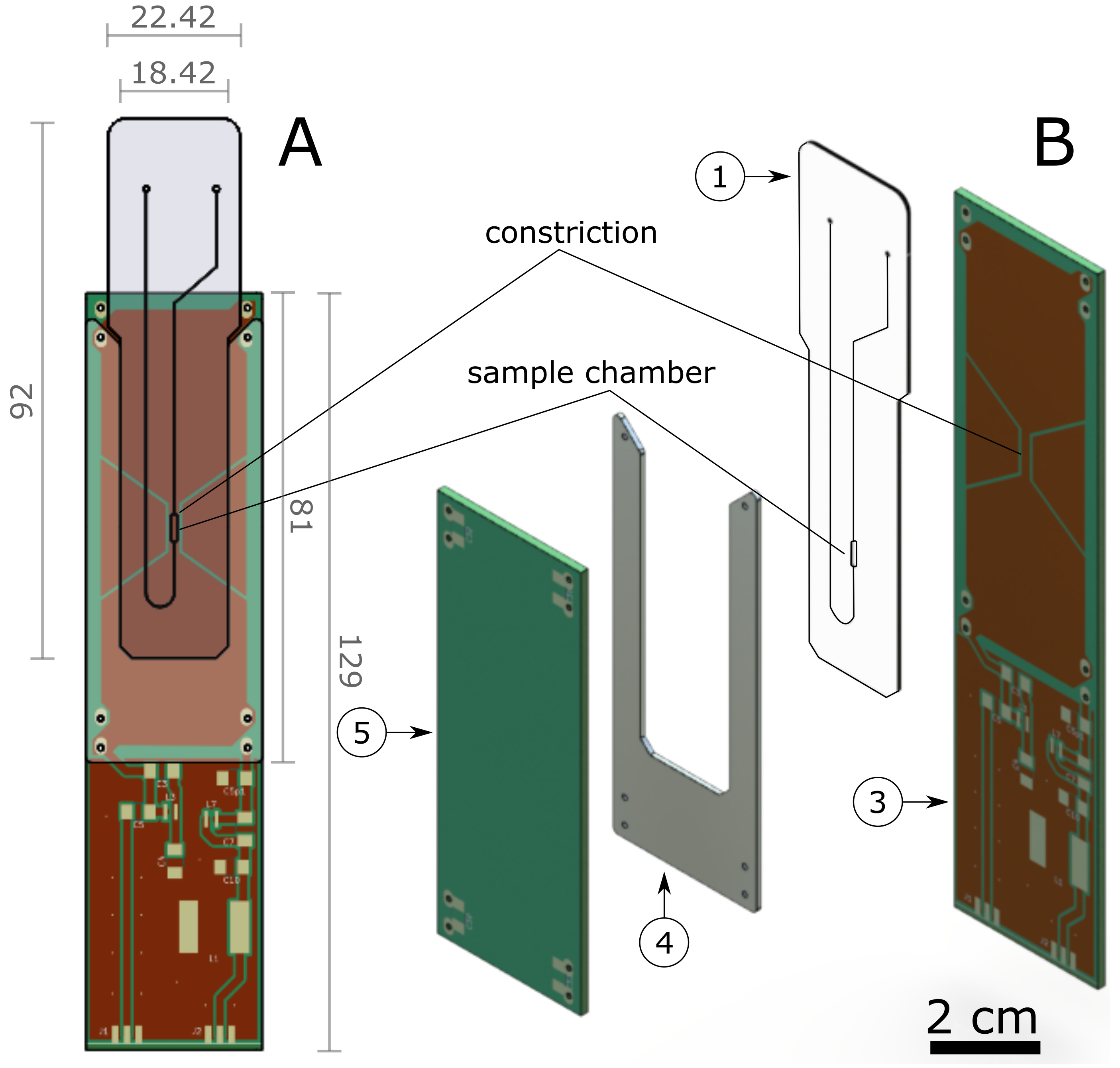}
\caption{Drawings of the detector assembly and the microfluidic device (1). A: front view (dimensions in mm); B: exploded view. Spacer (4) ensures the alignment of the sample chamber with the constrictions on the PCB planes. In A, PCB plane 5 is hidden to show the orientation of 1 with respect to PCB plane 3. Thickness of each of the PCB planes is 1.52 mm and the copper layers on the PCBs is 35 $\mu$m. Both the microfluidic device and the spacer are made from PMMA and have thickness of 0.9 mm and 1 mm respectively.}
\label{fig:rf-device-assembly}
\end{figure}

In this paper we present a novel modular probe designed for generic
microfluidic NMR experiments. The design is based on a transmission line
detector~\cite{stripline_jan,gream_2016} using inexpensive
printed circuit board (PCB) technology. A modular design ensures that detectors of
different geometry and size can be easily exchanged.
The NMR detector (including the tuning and matching circuit) is connected to the probe
skeleton through detachable connectors. Modularity of the probe not only allows
fast optimisation of the detector but also facilitates the use of dedicated detectors
for different microfluidic devices on a single probe base.
Once the probe base is in place, additional detectors can be manufactured
within 2-3 days for less than $\sim$200 $\geneuro$.
The probe is optimised for proton detected double resonance, with the
second channel tunable to different nuclei such as $^{13}$C, $^{15}$N, or
$^{31}$P. The probe skeleton is made almost entirely from easily obtainable raw materials
such as aluminium profiles of standard size, with only a small number of
custom-machined parts holding the structure together.
Detailed construction drawings and a complete set of CAD files is
available for download at \url{https://github.com/marcel-utz/modular-microfluidic-probe}.
The planar geometry of the detector accommodates exchangeable microfluidic devices
which can be designed freely for specific experimental requirements as long as the
overall shape of the device is maintained. \fig{fig:rf-device-assembly} shows the microfluidic device and the detector with their relative positioning.

In the remainder of this paper, we discuss the probe design in more detail,
and characterise its performance in terms of sensitivity and resolution.
We also demonstrate its application in microfluidic cell culture,
protein NMR spectroscopy, and micro-imaging.
\section{Probe Design}
\label{sec:probe-design}
\cbstart\
\cbend\
The probe assembly design is based on readily and easily available materials and parts
with minimal need for customised fabrication.
All mechanical and electronic components are readily available from commercial sources,
with the exception of a small number of precision-machined parts which are used
to hold the assembly together. These can be ordered from commercial prototyping
manufacturers or machined in-house if mechanical workshop facilities are available.

\begin{figure}
\centering
\includegraphics[width=.8\linewidth,keepaspectratio=true]{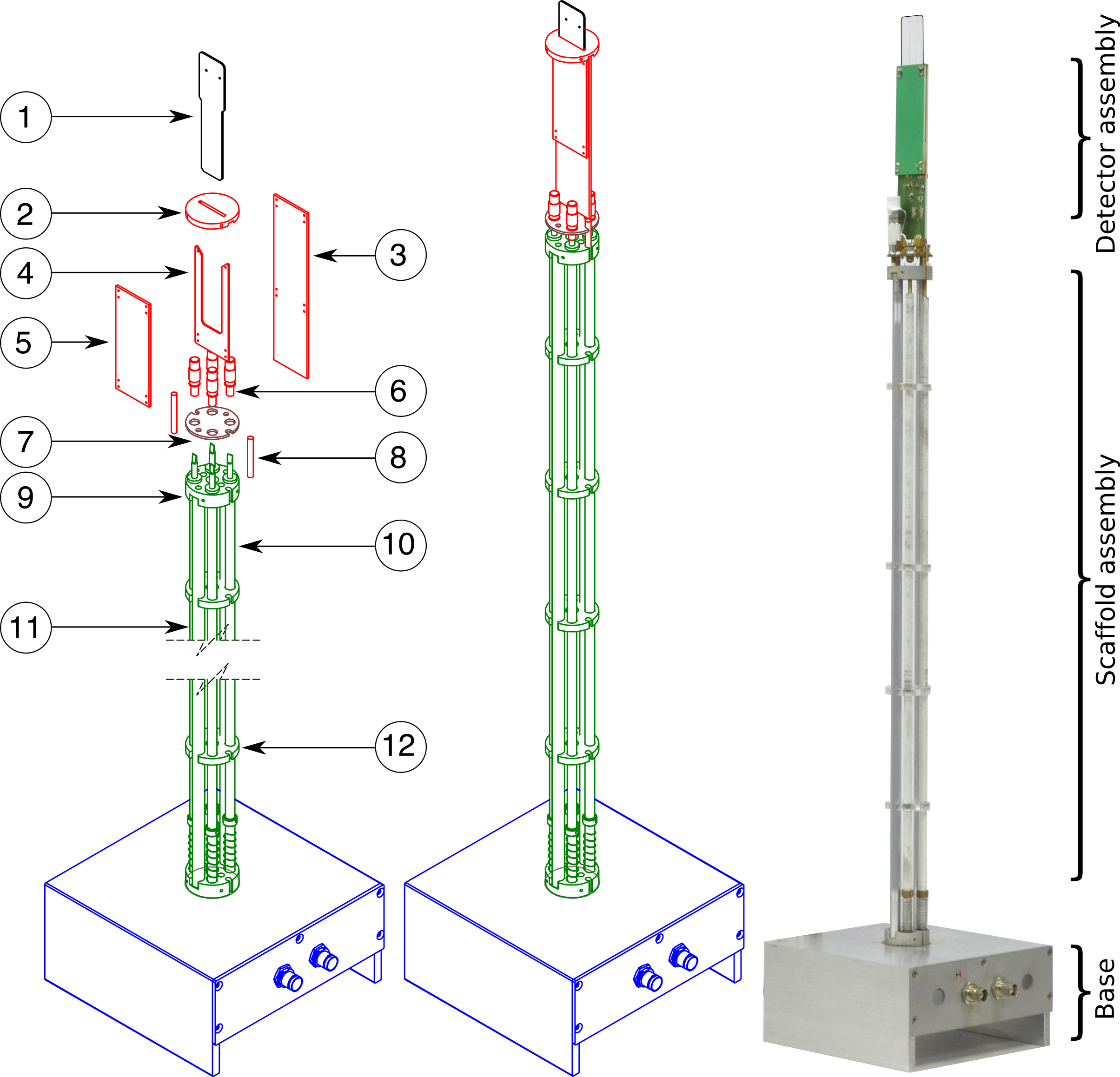}
\caption{Exploded view of the probe design (left) and assembled probe (center).
Probe base (blue) and scaffold assembly (green) with TMPCB (brown, 7) provides a
generic mounting point for the NMR detector assembly (red).
A microfluidic device (1) can be inserted from the top of the probe.
The device is held in place by a spacer (4) between the detector planes (3 and 5)
ensuring alignment of the sample chamber with the sensitive area on the PCB planes.
The modular NMR detector is made up of parts 3,4 and, 5 and is
connected to the TMPCB (7) through standard coaxial (SMP) connectors.
The scaffold assembly is made up of aluminium tubes (11), scaffold support disks (9)
and, auxiliary support disks (12).}
\label{fig:probe-explode}
\end{figure}

The probe is composed of three main assemblies: A) probe base; B) scaffold assembly;
C) detector assembly (\fig{fig:probe-explode}).
The  base (blue in \fig{fig:probe-explode}) and the scaffold assembly
(green in \fig{fig:probe-explode}) together make the skeleton of the probe.
Both the base and scaffold assembly are made from standard aluminium profiles.
The base contains the BNC connectors for RF (Radio Frequency) input.
These BNC connectors are connected to semi-rigid coaxial cables for
RF transport in the probe.

\begin{figure}
\centering
\includegraphics[width=.5\linewidth,keepaspectratio=true]{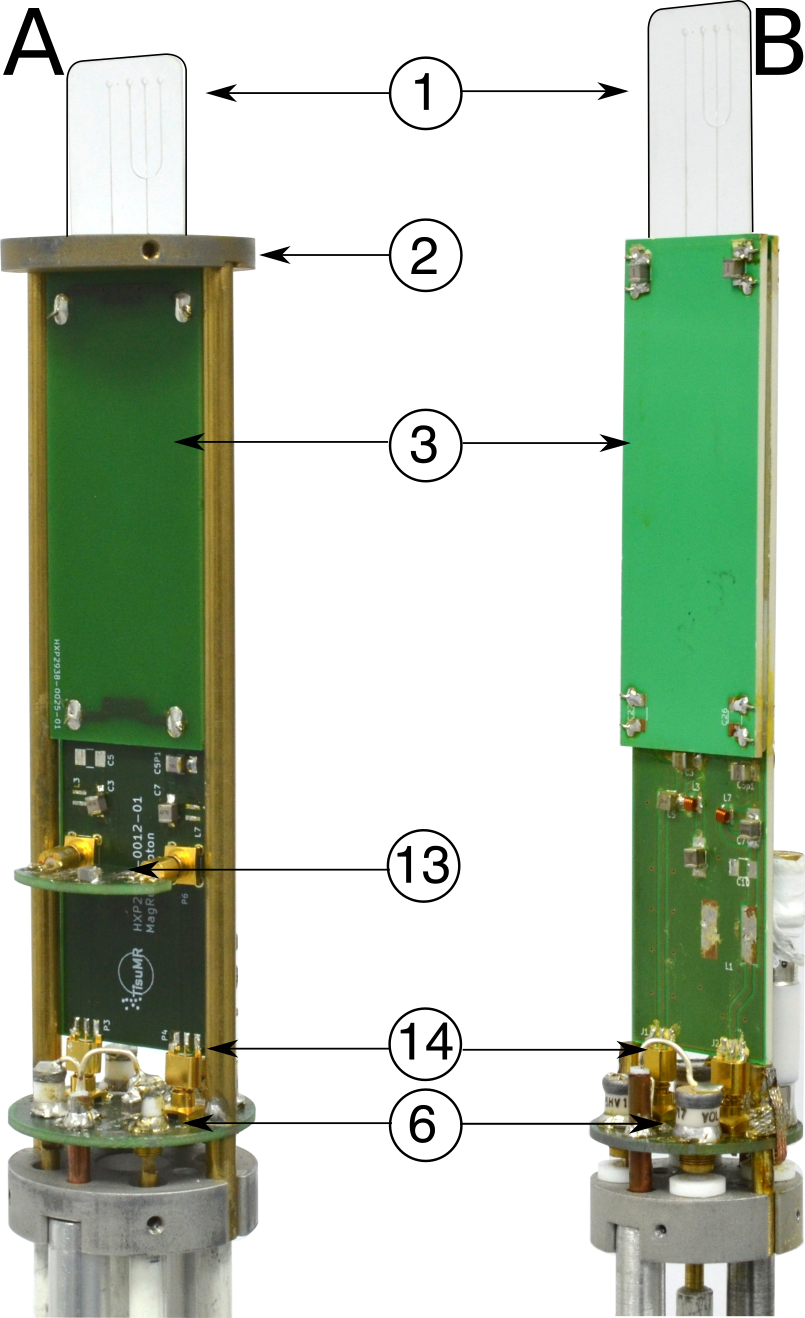}
\caption{Dual channel probes with (A) and without (B) modular RF insert (13) for X channel.
O.D. of A is 44.45 mm and B is 38 mm. Circuit elements of X channel are integrated on the
PCB in B making it compact. Detachable SMP connectors (14) are used to
connect the RF insert (13) and the detector assembly.}
\label{fig:ProbePhoto}
\end{figure}

\begin{figure}
\centering
\includegraphics[width=\linewidth,keepaspectratio=true]{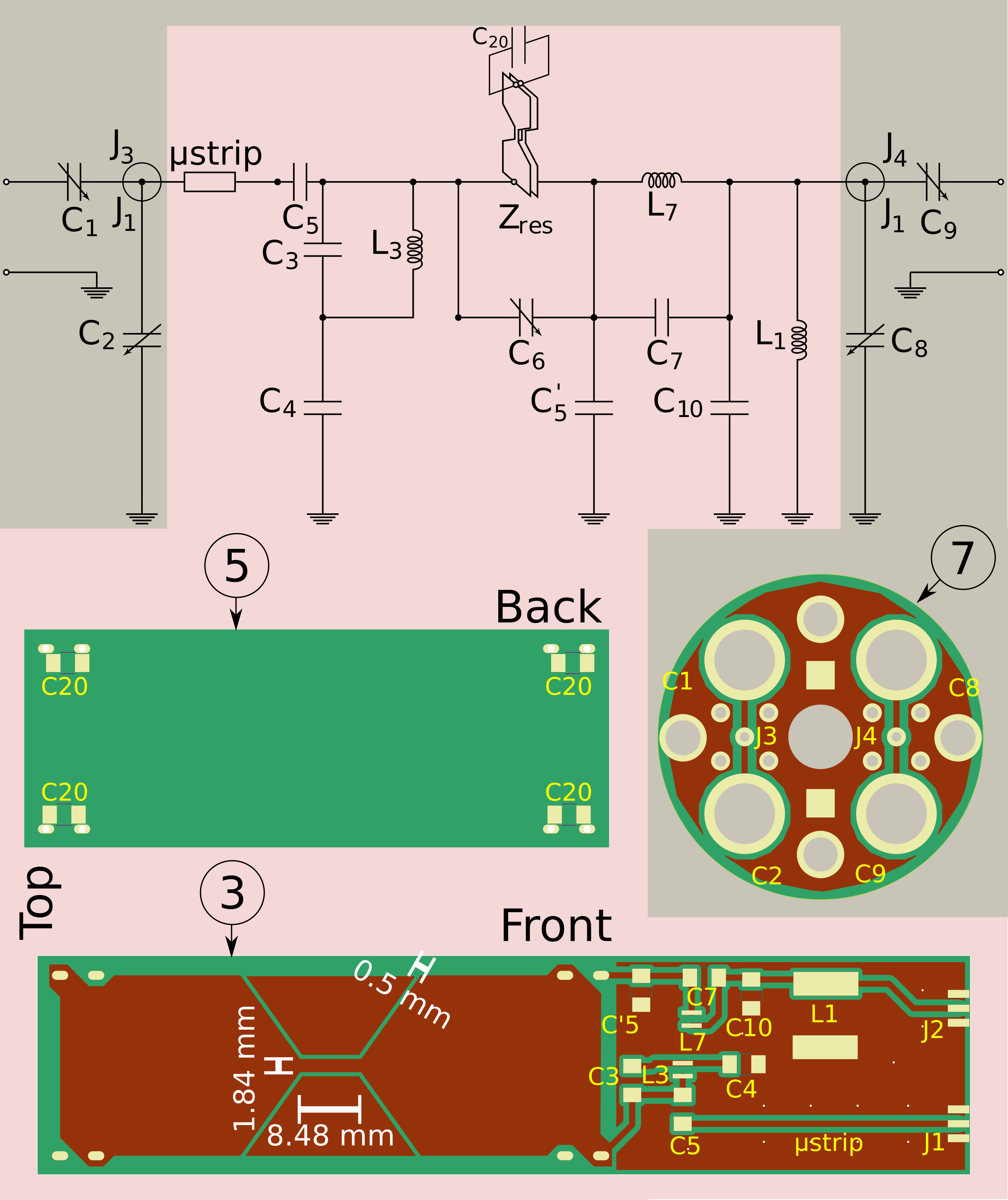}
\caption{Circuit diagram of the double resonance probe. The background colour relates the different circuit elements on different PCBs. The back side of the PCB plane 5 and the front side of the PCB plane 3 is shown in horizontal orientation. The front side of the TMPCB 7 is shown in vertical orientation. Trimmer capacitor $C_6$ is placed on the back side of PCB plane 3. The copper layer is shown in brown and the solder pads to place the circuit elements are shown in off-white. An insulating solder mask layer over the copper layer prevents accidental short-circuiting.
}
\label{fig:circuit}
\end{figure}

The scaffold assembly is made up of two hollow aluminium tubes (11) held together with
precision machined parts at both ends (scaffold support disks (9)).\cbstart\ The semi-rigid cables (RG 402, 50~$\Omega$) are  threaded through these hollow aluminium tubes
to reach the detector assembly.\cbend\
Auxiliary support disks (12) provide additional rigidity for the scaffold assembly.
In our case, these were laser cut from 5~mm thick poly(methyl methacrylate) (PMMA) sheets,
but they may also be made by conventional machining.
A tuning and matching PCB (TMPCB, 7) is held in place on the top of the probe skeleton by
the detector support rods (8), with the semi-rigid coaxial cables providing
additional support.
The TMPCB is permanently fixed to the probe base, and serves as attachment point
for the exchangeable detector assembly.\cbstart\ It also provides
trimmer capacitors (6) for secondary tuning and matching of the detector.\cbend\ The detector assembly is connected to the TMPCB through a pair of
low-loss SMP connectors (14 in \fig{fig:ProbePhoto}). \cbstart\ The SMP connector receptacle male pin ($J_3$ and $J_4$ in \fig{fig:circuit}) is soldered on the TMPCB and the SMP connector jack male pin is soldered on the PCB plane 3 ($J_1$ and $J_2$ in \fig{fig:circuit}). The TMPCB and 3 are connected by a coaxial connector SMP plug adapter.\cbend\ While each detector assembly is roughly
tuned and matched with fixed circuit elements soldered onto it,
the adjustable capacitors on the TMPCB are used for
fine tuning and matching
at the magnet through tuning rods (10) which reach all the way to the probe base.

The base, the scaffold assembly and the tuning and matching PCB all have a free circular
axial bore providing access from the bottom of the probe, which can be used for
temperature and environmental control and/or fluidic connections to the sample as
needed. The detector assembly is made up of two PCBs (3 and 5) and a spacer (4) between them.\cbstart\ The spacer is made from a 1 mm thick PMMA sheet.\cbend\ The detector used here is based on a planar transmission line geometry as described elsewhere~\cite{gream_2016,stripline_jan}; other structures such as spiral
coils and Helmholtz pairs are also possible, and can easily be realised on a PCB.
The transmission line detector consists of a pair of copper
planes, each with a constriction at the centre.  This geometry gives rise to an
electromagnetic eigenmode with a strong anti-node of the magnetic field between
the two planes at the site of the constriction. This concentrates the rf field
and the detection sensitivity
onto the sample area.
The shape and size of the copper structures have been chosen such as to produce
an eigenfrequency of this mode around 630~MHz. The PCBs have been designed with
solder points at the back side for chip capacitors; these can be used to tune the
resonance down to the desired proton Larmor frequency (500 or 600~MHz in the present case).
Both the PCBs (3 and 5) also carry circuitry for channel separation and primary
matching of the detector. A copper or brass auxiliary disk (2) holds the PCBs
together at the top, and forms the top end of the probe assembly.
The probe sheath (not shown in~\fig{fig:probe-explode}) is fixed to the auxiliary disk (2)
and the scaffold assembly with screws.
An exchangeable microfluidic device (1) can be inserted from the top of the detector.
Microfluidic devices can be used as passive sample holders, filled before putting them
into the probe,
or the device can be actively perfused and operated in-situ.\cbstart\ The device inlet and outlet are sealed by an optical adhesive film (MicroAmp, AB Applied Biosystems)
when using the device as passive sample holder.\cbend\ The microfluidic device is held in place by the spacer
between the two PCB planes coinciding the sample chamber with the constriction on the
detector. The 45-degree edges of the device outline help with precise and reliable
alignment.
Since the detector assembly is attached to the probe base through SMP
connectors, it can easily be exchanged.
Using standard PCB processes to manufacture the detector provides great flexibility
to optimise detector performance for different applications, including different
microfluidic sample sizes, and different combinations of target nuclei.

\fig{fig:ProbePhoto} shows two different versions of the probe with and without the RF insert (13)
for the X channel. In (A) the X channel can be tuned to different nuclei by exchanging (13).
However, in (B) the circuitry of the RF insert (13) is integrated on the PCB plane (5),
making the design more compact.
 Probe (A) is compatible with a Varian 600 Premium shielded 14.1 T magnet, with an outer diameter (O.D.) of 44.45 mm including the sheath. Probe (B), with an
O.D. of 38 mm including the sheath, is designed to fit inside a
Bruker narrow bore shim stack.
Probe (B) also fits inside a Bruker micro-imaging gradient unit. Naturally, B can also be used with
wide bore magnets with appropriate spacers to keep the probe centered.

\begin{figure}
\centering
\includegraphics[width=\linewidth,keepaspectratio=true]{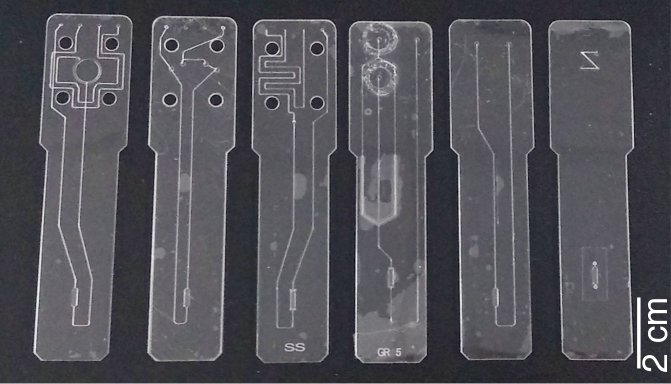}
\caption{Microfluidic devices for different applications compatible with the current setup.
From left: Devices for perfusion of tissue slice on a chip, peristaltic pumping to induce flow,
hydrogenation on a chip, droplet generator, simple design to fill the sample chamber and,
to grow cells on the device.}
\label{fig:device}
\end{figure}

The microfluidic devices were made from inexpensive PMMA sheets by laser cutting and
thermal bonding, as described in detail elsewhere \cite{yilmaz_bonding}.\cbstart\ PMMA was chosen for ease of fabrication, but also due to the close match in magnetic susceptibility with that of water (Table~\ref{tab:susceptibility}).\cbend\
Briefly, the devices consist of a 0.5~mm thick middle layer between a  top and
bottom layer of 0.2~mm thickness each. The sample chamber and fluidic network
are cut into the middle layer. The fluidic network on the device can be designed
freely as long as the overall shape of the device is preserved, and the location
of the sample chamber is maintained. The volume of the sample chamber is 2.5~$\mu$l which can be increased or decreased according to the experimental requirements. \cbstart \fig{fig:device} shows various microfluidic devices.\cbend
\cbstart
\begin{table}[h]
\centering
\caption{Magnetic susceptibility values of some relevant materials.\cite{wapler-2014,schenck-1996}}
\label{tab:susceptibility}
\begin{tabular}{||p{.3\linewidth}|p{.3\linewidth}||}
\hline
Material   & $\chi$/ppm	\\
\hline
PMMA		&-9.054	\\
Copper	&-9.63	\\
FR4	&-3.743		\\
Aluminium	&20.7	\\
Water & -9.02\\
\hline
\end{tabular}
\end{table}
\cbend
\cbstart
\cbend

\cbstart The double resonance circuit is based on the design shown in \fig{fig:circuit}.\cbend
 The transmission line detector is tuned to the desired proton
Larmor frequency by chip capacitors ($C_{20}$) located at its four corners,
as well as a non-magnetic trimmer capacitor ($C_6$). Two primary
matching capacitors ($C_5$ and $C_5'$) bring the impedance into the
vicinity of 50$\Omega$. A short microstrip
with a specific impedance of 50$\Omega$
on the PCB then provides a connection to the secondary tuning and matching elements on the TMPCB ($C_1$ and $C_2$).
The lower frequency channel is connected to the detector through a high-frequency band reject ($C_7$, $L_7$).
Tuning is achieved by $L_1$ and $C_4$ in combination with
the complex admittance of the detector at the X channel frequency.
To maximise efficiency, $L_1$ should have as high a $Q$-factor as possible. We have
used commercially available flat copper coils of 5 mm diameter and 4.5 turns in the case of
\textsuperscript{15}N at 60 MHz. Tuning to \textsuperscript{13}C, either at 125 or at 150 MHz
could be achieved without $L_1$. \cbstart Details of all the circuit elements are given in table~\ref{tab:circuit}.
\begin{table}[h]
\centering
\caption{Details of the circuit elements to tune the probe at 600 MHz and 60 MHz.}
\label{tab:circuit}
\begin{tabular}{||p{.3\linewidth}|p{.2\linewidth}|p{.3\linewidth}||}
\hline
Circuit element   & Value	&Manufacturer Part Number	\\
\hline
$C_1$, $C_2$ and, $C_8$		&1-23 pF	& NMAM25HV\\
$C_9$	&3-23 pF	& NMNT23-6\\
$C_6$	&3-23 pF	& NMAJ25HV\\
$C_{20}$  & 2.4 pF		& S111TUE\\
$C_3$, $C_7$	&4.3 pF		& S111TUE\\
$C_4$	&82 pF	& S111TUE\\
$C_5$, $C^{'}_{5}$	&2.4 pF	& S111TUE\\
$L_1$	&111 nH	& 1010VS-111ME\\
$L_3$, $L_7$	 &16.6 nH	& 17NGLB\\
$J_1$, $J_2$ & -&0734153591\\
$J_3$, $J_4$ & -&SMP-MSLD-PCT-5\\
\hline
\end{tabular}
\end{table}
\cbend
\cbstart
\cbend
\section{Materials and Methods}
All the parts and the probe assembly were designed in Solidworks (Solidworks,
Dassault Syst\`{e}mes). Prefabricated profiles made from 6061-T3 aluminium
(Aluminum Warehouse, UK) were used for the mechanical structure of the probe
(base platform, scaffold tubes, sleeve). A few customised precision-machined
parts (scaffold support disks, top disk) were obtained from ProtoLabs (Telford,
UK) after providing CAD files. A small number parts were machined in-house
(detector support rods, auxiliary support disks). All electronic components
including semi-rigid transmission lines were purchased from DigiKey Ltd\cbstart\ and Farnell element14, UK. Sodium acetate and $^{15}$N labeled urea were purchased from Sigma-Aldrich, UK and Dulbecco Modified Eagle's Medium (DMEM) supplemented was purchased from Life Technologies, USA.\cbend\ Cell cast PMMA sheets for microfluidic devices were bought from Weatherall Equipment
(Wendover, UK). Microfluidic devices were designed on AutoCAD. PMMA sheets were
cut to desired size and functionality (holes or channels) using a laser
engraving system  (L3040 from HPC Laser LTD, Elland, UK), and then bonded using the protocol
described in~\cite{yilmaz_bonding}.

The precise dimensions of the detector were optimised for a given sample volume
using finite element calculations in COMSOL, as described in \cite{gream_2016}.
The RF response of the probe circuit was simulated using a symbolic network
analysis code~\cite{gream-thesis} in order to estimate appropriate component
values. Final tuning was done by trial and error using an Agilent FieldFox 4 GHz
network analyser. PCB layouts were designed using KiCAD, and then sent to
beta-layout GmbH (Aarbergen, Germany) and P.W. Circuits Ltd.~(Wigston, UK) for
manufacturing. Easily and inexpensively available FR4 PCB material
(\fig{fig:ProbePhoto}(A)) proved adequate for some purposes, but limited
resolution to about 10 Hz. A high-quality Teflon-based RF PCB material (RO3035,
Rogers Corp., USA) was therefore used for optimal performance in both
sensitivity and resolution (\fig{fig:ProbePhoto}(B)). The probe skeleton was
assembled from prefabricated aluminium tubes held together with the precision-machined
scaffold disks using brass screws. Semi-rigid transmission lines were then
threaded through the tubes, and the TMPCB plate was soldered on top. The
detector planes (3 and 5) and the PMMA spacer between them were assembled using
short bits of 0.5~mm\cbstart\ diameter copper wire bought from RS components, UK,\cbend\ which were soldered to the appropriate pads in the four corners of the detector structure.

\cbstart
\subsection{Details of NMR and MRI measurements}
NMR spectra were obtained on a Varian VNMRS console attached to a Varian 600 premium shielded 14.1 T magnet. Field maps in the \fig{fig:field map} were created by gradient echo (FLASH) pulse sequence with echo times ($t_{e}$) of 6 and 16 ms and repetition time ($t_{r}$) of 100 ms on a sample of 130 mM  sodium acetate dissolved in H$_2$O. The peak shown in the \fig{fig:1H-nutation} is of H$_2$O in an aqueous solution of 100~mM $^{15}$N-urea. \fig{fig:lineshape} shows proton spectrum of 2.5 $\mu$l of 130 mM  sodium acetate dissolved in H$_2$O. The spectrum was recorded after pre-saturating the water signal at a nutation frequency of 100~Hz. 32 transients were averaged with a repetition delay of 3~s. The proton spectrum of DMEM (Dulbecco
modified Eagle's medium) cell growth medium in \fig{fig:media-spec} was acquired with 256 scans in 20 mins.
The left panel of \fig{fig:HSQC} shows a $^{13}$C-$^{1}$H heteronuclear single
quantum correlation (HSQC) spectrum, acquired in 12 mins (32 t$_1$ increments
and 16 scans) from a 2.5 $\mu$l sample of 100 mM $^{13}$C-glucose dissolved in
water. The right panel shows a $^{15}$N-$^{1}$H-HSQC spectrum, acquired in 400
mins (64 t$_1$ increments and 128 scans) from 2.5 $\mu$l sample of 1 mM
$^{15}$N-ubiquitin (17 $\mu$g) dissolved in 10 mM phosphate buffer at pH 7.

For MRI experiments precision-cut murine liver slices (PCLS) were obtained by first extracting a
tissue core of 8 mm diameter using a biopsy punch. PCLS of 250-300 $\mu$m
thickness were cut from the core using a Krumdieck Tissue slicer, and then
reduced to 3~mm diameter using a smaller punch. The slices were incubated in
well-plates at 37$^\circ$C and 80\% O\textsubscript{2}, 5\% CO\textsubscript{2}
for 2 hours before imaging. Imaging experiments were performed on a Bruker
AVANCE III spectrometer on a Bruker Active Shield II wide bore 11.7 T magnet equipped with a Bruker Micro2.5 gradient system. The image shown in \fig{fig:tisli}(B) was acquired using a FLASH pulse sequence with $t_R=120$~ms and $t_E=15.3$~ms. The slice thickness of  1~mm encompasses the entire thickness of the tissue. \fig{fig:tisli}(C) shows a
RARE image using the same parameters, with 8 echos acquired per scan, thus
reducing the acquisition time to 8 minutes. A spin echo image is shown in
\fig{fig:tisli}(D), acquired over 25~min. Finally, \fig{fig:tisli}(E) shows a
$B_0$ field map, which has been obtained by acquiring two separate FLASH images
with echo times $t_E=5.3$~ms and 15.3~ms, respectively.
\cbend

\section{Results and Discussion}

\begin{figure} \centering
\includegraphics[width=\linewidth,keepaspectratio=true]{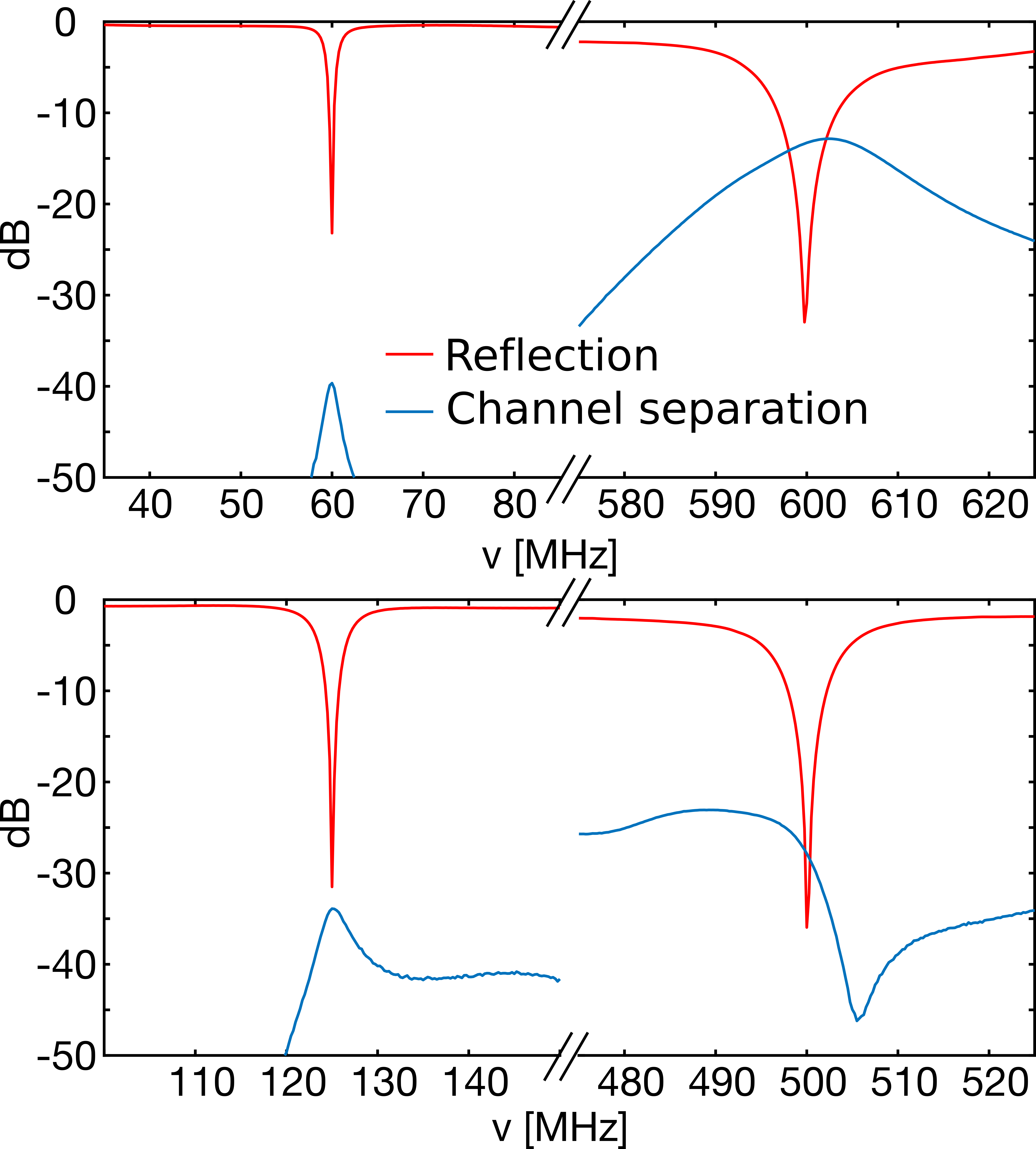}
\caption{ Experimental tuning and matching curves measured at 600 MHz, 500 MHz,
60 MHz (\textsuperscript{15}N frequency at 14.1 T) and 125 MHz
(\textsuperscript{13}C frequency at 11.7 T). Probe can be tuned to 2
frequencies at a time, above results were measured in pairs of 600 MHz with 60
MHz and 500 MHz with 125 MHz. Decoupling can be performed at all the frequencies
as the channel separation is better than 15 decibels at all frequencies.}
\label{fig:tandm}
\end{figure}

The probe is designed for double resonance, with proton-detected experiments in
mind. The circuit is optimised for efficiency and sensitivity at the
high-frequency channel, while allowing RF pulses to be applied at the X
frequency as well. As already explained, the X channel tuning frequency can be
changed by simply exchanging either only a small RF-insert or the complete
detector assembly. \fig{fig:tandm} shows the measured reflection and
transmission scattering coefficients ($S$-parameters) as a function of frequency
for two combinations of proton (600 MHz, 500 MHz) and X channel frequencies (60
MHz and 125 MHz), corresponding to the Larmor frequencies of $^{15}$N and
$^{13}$C at magnetic fields of 14.1 T and 11.7 T respectively. The measurements
were performed in the pairs of 600 MHz and 60 MHz and 500 MHz and 125 MHz. The
loaded $Q$ factors, determined from the -3dB points of the reflection curves,
are 42, 49, 32, 36 at 600 MHz, 500 MHz, 125 MHz and, 60 MHz respectively. The
channel separation is better than -15dB at all operating frequencies.

\begin{figure} \centering
\includegraphics[width=\linewidth,keepaspectratio=true]{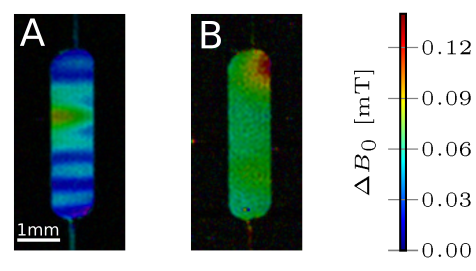}
\caption{B\textsubscript{0} field map of the sample chamber for FR4 (A) and
Rogers RO3035 (B) PCB material acquired at 11.7 T. Both the field maps were constructed from 2
gradient echo images acquired on a chip filled with 130 mM sodium acetate
dissolved in H\textsubscript{2}O through same parameters and pulse sequence.}
\label{fig:field map}
\end{figure}

\cbstart
\cbend
Initially, FR4 PCB material was used for the
detector assembly because of its low cost and easy availability. Inhomogeneities
in the B$_0$ field due to the glass fibre rovings in FR4's composite backing can
be seen as fringes in~\fig{fig:field map} (A). \cbdelete B$_{0}$ field maps of the sample
chamber were constructed from 2 gradient echo (FLASH) images for both FR4
and RO3035 PCB materials with the same chip~(\fig{fig:field map}). The
brightness of each pixel is proportional to the sum of the magnitude of the two
images, whereas the colour encode the phase difference between them. The full
colour range from red to blue spans phase differences from $-\pi$ to $+\pi$.
\cbdelete
B$_{0}$ field map of RO3035 PCB material clearly show better B$_{0}$ field homogeneity compare to FR4. To avoid the inhomogeneities in FR4 and therefore to achieve better resolution,
detectors were made from RO3035. While RO3035 provides better
resolution, it is more expensive and takes longer to obtain compared to FR4. It
therefore makes sense to first prototype new detector designs in FR4, and only
use the more expensive RO3035 once the design has been validated.

\begin{figure}
\centering
\includegraphics[width=.95\linewidth,keepaspectratio=true]{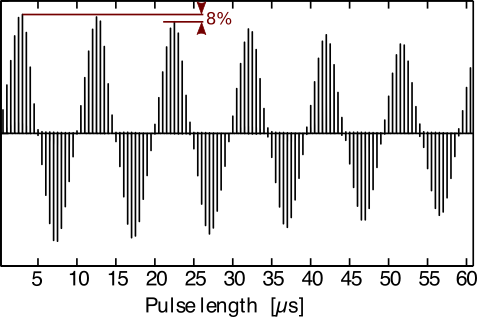}
\caption{$^1$H nutation acquired for water peak from a sample of 100 mM $^{15}$N-urea dissolved in water measured at 600 MHz. The 810$^{\circ}$/90$^{\circ}$ ratio is 92\% at 600~MHz. The 90$^{\circ}$ pulse length is below 2.5 $\mu$ sec corresponding to 100 kHz radio-frequency field for an input power of 100 W. Similar B$_{1}$ efficiency and homogeneity was observed at 500 MHz.}
\label{fig:1H-nutation}
\end{figure}

\cbstart
\cbend
\cbdelete
The RF field at proton frequency shows excellent homogeneity, with a
810$^{\circ}$/90$^{\circ}$ ratio of more than 90\% measured from $^1$H nutation at 600~MHz shown in~\fig{fig:1H-nutation}. The
probe generates 100~kHz nutation frequency from 100~W input power both at
600~MHz and 500~MHz, corresponding to RF field efficiency ($B_{1}/\sqrt{power}$)
of 10 kHz/$\sqrt{W}$. The detector geometry has been optimised for a sample
volume of 5~$\mu$l. It compares reasonably well to the value of 22
kHz/$\sqrt{W}$ that has been reported by Finch et al.~\cite{gream_2016} for
single-frequency detector of half this size at a Larmor frequency of 300~MHz.
For the X channel, 100~W of input power produces nutation frequencies of 6 kHz for
$^{15}$N and 14 kHz for $^{13}$C. No signs of arcing were obseved at this power
level for pulses up to 1~ms duration.

\cbstart
\cbend


\begin{figure}
\centering
\includegraphics[width=.80\linewidth,keepaspectratio=true]{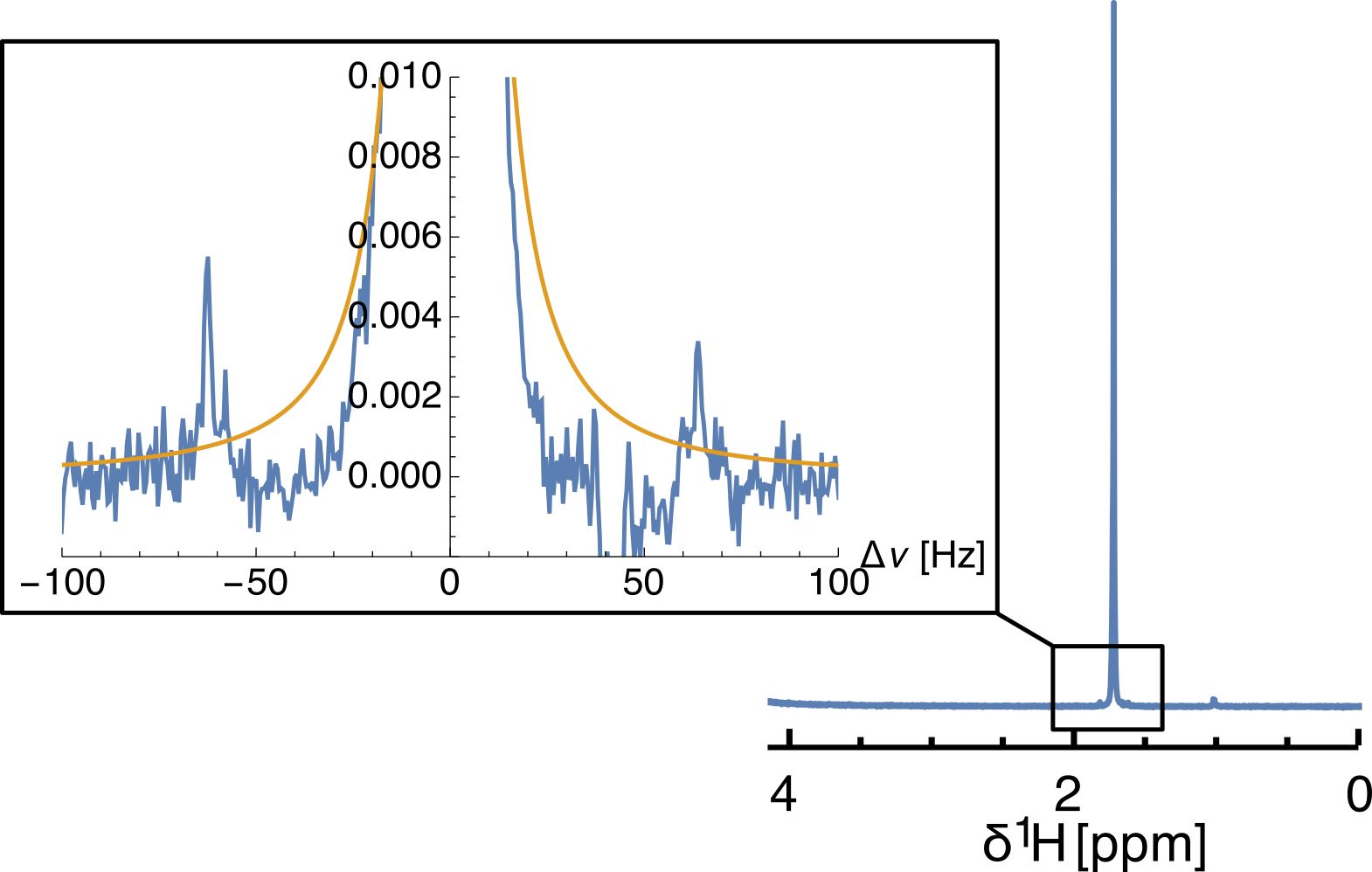}
\caption{Spectrum of 130 mM sodium acetate dissolved in water acquired in 32 scans with water presaturation. Carbon satellites can be seen in the expanded base of the acetate line. The orange curve shows a lorentzian lineshape with full width at half maximum of 3.35 Hz.}
\label{fig:lineshape}
\end{figure}

\begin{figure}
\centering
\includegraphics[width=\linewidth,keepaspectratio=true]{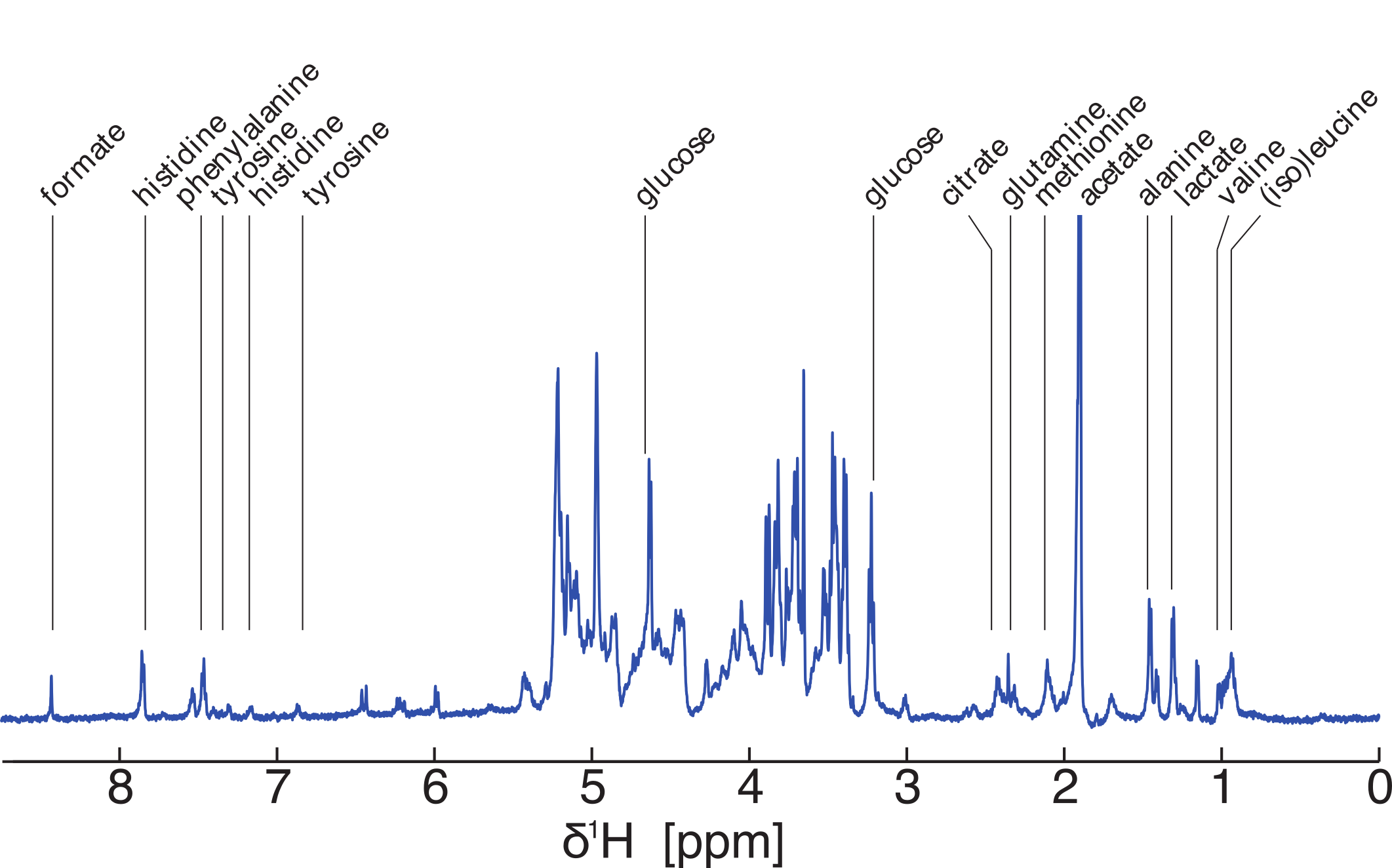}
\caption{Spectrum of culture growth media recorded in 256 scans with water presaturation. The total experimental time was 20 min for 2.5 $\mu$l sample volume.}
\label{fig:media-spec}
\end{figure}
\cbdelete
\fig{fig:lineshape} shows the proton spectrum of sodium acetate. The inset shows an expanded view of the base of the acetate peak. Carbon satellites can be seen well above the noise level. The orange curve
corresponds to a Lorentzian lineshape with full width at half maximum of 3.35
Hz, corresponding to 0.0056 ppm. This is very similar to the resolution reported
by Finch et al. at 300~MHz. While the residual cause of broadening is still
under investigation, we have reason to suspect imperfections in the fabrications
of the PMMA microfluidic devices to play a major role. Efforts to mitigate these
and to achieve more highly resolved proton spectra are currently underway in our
laboratory and will be reported at a later occasion.  The limit of detection
$nLOD_{\omega}=\frac{3n\sqrt{{\Delta}t}}{SNR}$ ($n$ is the number of spins and
$\Delta t$ is 1/linewidth of the observed resonance) is 1.4 nmol s$^{1/2}$ for
protons at 600 MHz, slightly better than the value reported for a smaller
detector by Finch et al.~\cite{gream_2016}.

\begin{figure} \centering
\includegraphics[width=\linewidth,keepaspectratio=true]{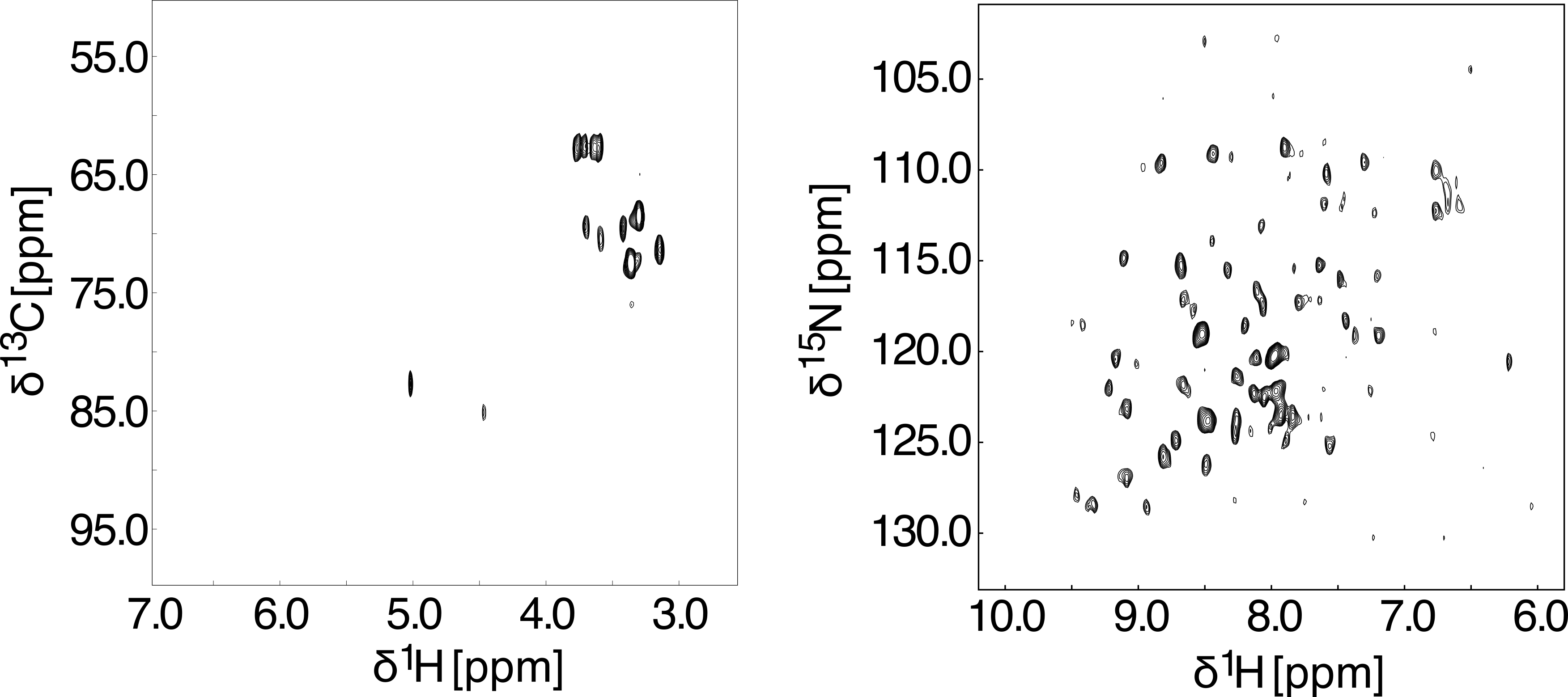}
\caption{ $^1$H-$^{13}$C HSQC spectrum of $^{13}$C labeled glucose (left) and
$^1$H-$^{15}$N HSQC of $^{15}$N labeled Ubiquitin (right) both acquired from
2.5$\mu$l sample volume at 14.1 T. Glucose HSQC spectrum was acquired in 12
minutes from a 100 mM sample. Ubiquitin HSQC spectrum was acquired in 400
minutes from a 1 mM sample.} \label{fig:HSQC} \end{figure}

Even though the resolution of the proton spectra still needs to be improved, it
is sufficient to analyse complex mixtures. \cbdelete The proton spectrum of DMEM (Dulbecco
modified Eagle's medium) shown in \fig{fig:media-spec} demonstrates the suitability of this
probe for metabolomic studies of microfluidic cultures of cells, tissues, and
small organisms.
\cbdelete
\fig{fig:HSQC} demonstrates proton-detected double resonance experiments in the
current setup. The left panel shows a $^{13}$C-$^{1}$H heteronuclear single
quantum correlation (HSQC) spectrum of $^{13}$C-glucose and the right panel shows a $^{15}$N-$^{1}$H-HSQC spectrum of $^{15}$N-ubiquitin (17 $\mu$g). All
expected cross peaks are present in the spectrum. The ability to obtain
meaningful  \textsuperscript{15}N-\textsuperscript{1}H correlation spectra from
such small samples could prove very useful in the context of proteins that are
difficult to express and therefore only available in small quantities. The
integration of protein NMR and microfluidic technology also opens new
possibilities for high-throughput kinetics and binding studies.

\begin{figure}
\centering
\includegraphics[width=\linewidth,keepaspectratio=true]{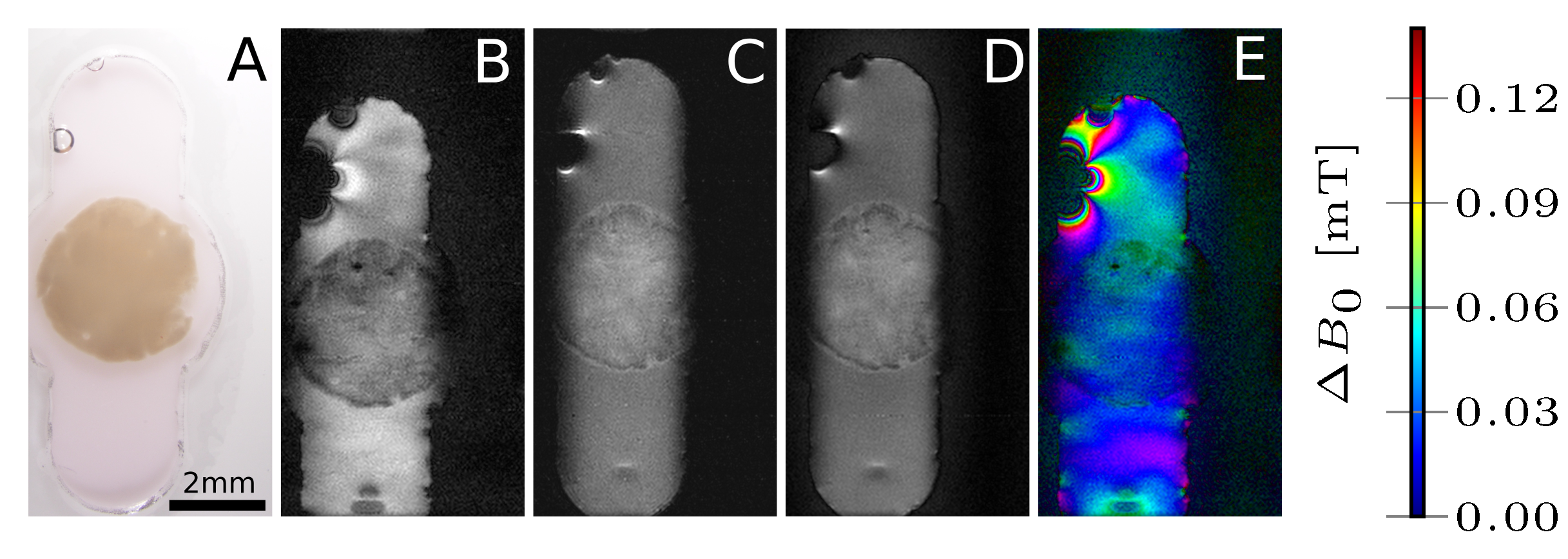}
\caption{MR images of a freshly cut liver slice from a mouse. A: optical micrograph; B: Gradient echo image acquired in 1 minute; C: RARE8 image acquired in 8 minutes; D: Spin echo image acquired in 30 minutes; E: magnetic field map.}
\label{fig:tisli}
\end{figure}

\cbdelete
The probe can also be used to acquire micro magnetic resonance (MR) images using
a Bruker micro-gradient unit. To this effect, the aluminium probe sleeve must be
replaced by an electrically insulating material in order to avoid image
distortions due to eddy currents in the sleeve.\cbstart\
A shorter aluminium tube with 3D printed top part (made from ABS plastic)
to match the probe height was used.\cbend\ \fig{fig:tisli} shows MR images
of mouse liver tissue slice immersed in DMEM cell growth medium. The slice was
placed in a microfluidic device with circular sample chamber centred at the
detector location, as shown in \fig{fig:tisli}(A). The image shown in
\fig{fig:tisli}(B) was acquired using a FLASH pulse sequence with; the lateral dimensions of all MR images in \fig{fig:tisli} are 512$\times$512 pixels at a resolution of 30~$\mu$m. \fig{fig:tisli}(C) shows a
RARE image using the same parameters. A spin echo image is shown in
\fig{fig:tisli}(D). Finally, \fig{fig:tisli}(E) shows a
$B_0$ field map, which has been obtained by acquiring two separate FLASH images.

\section{Conclusions}
We have presented a novel modular NMR probe assembly for generic high
sensitivity microfluidic NMR experiments.
The probe is built from the readily accessible materials, using only a minimum
of custom-made parts.
The modularity of the probe enables to use different detectors
tailored for specific experimental requirements on the same probe skeleton.
The detector itself is made from easily available PCB materials.
The use of PCB technology offers great flexibility, and allows to modify the
probe circuit and design without affecting the probe geometry.
The probe has excellent B$_1$ homogeneity and its sensitivity is in line with
the best performances that have been reported in this range of probe volumes.

Using simple microfluidic devices made from PMMA sheets by laser cutting,
spectral resolution around 0.005~ppm is achieved in proton spectra at
600~MHz. This is sufficient for primary metabolic studies, but some
applications could benefit from better resolution. The probe has already been
used for various  experiments including reaction monitoring~\cite{Fang-2018},
micro-imaging of mouse liver tissue slices, and metabolic studies of
mammalian cells. The probe is designed for double resonance, and
acquisition of proton-detected heteronuclear correlation experiments
has been demonstrated for small molecules as well as proteins.
In the future we intend to add a field-frequency lock channel,
as well as additional functionality for advanced microfluidic perfusion with temperature and flow control.
\section{Acknowledgement}
This research was supported by the "TISuMR" project,
funded through "Future and Emerging Technologies" (FETOPEN)
call of the EU Horizon 2020 research framework. The authors would like to
thank Mr.~William Hale for help with
making microfluidic devices, Dr.~Katrin Deinhardt for providing
the mouse liver tissue, Dr.~Bishnubrata Patra for tissue slicing and Dr.~J\"{o}rn Werner for providing the
labelled ubiquitin sample and for his support in the acquisition
and processing of the HSQC spectrum.
\clearpage
\bibliographystyle{unsrt}
\bibliography{science}
\end{document}